\documentclass[aps, pre,
 amsmath,amssymb,
 reprint,%
]{revtex4-1}
\usepackage{changes}
\usepackage{dcolumn}% Align table columns on decimal point
\usepackage{bm}% bold math
\usepackage[utf8]{inputenc}
\usepackage[T1]{fontenc}
\usepackage{mathptmx}
\usepackage{etoolbox}
\usepackage{graphicx} %package for adding images.
\graphicspath{}
\usepackage{tabularx} %for table
\usepackage{paralist}%in para list
\usepackage{amsmath} %math
\makeatletter
\def\@email#1#2{%
 \endgroup
 \patchcmd{\titleblock@produce}
  {\frontmatter@RRAPformat}
  {\frontmatter@RRAPformat{\produce@RRAP{*#1\href{mailto:#2}{#2}}}\frontmatter@RRAPformat}
  {}{}
}%
\makeatother
\begin{document}
\preprint{AIP/123-QED} 
%\begin{document}
\preprint{AIP/123-QED} 
\title{\bfseries Coupling of Lipid Phase Behavior and Protein Oligomerization in a Lattice Model of Raft Membranes}
%\runningtitle{Lipid Regulation of Protein Oligomers}
\author{Subhadip Basu}

\affiliation{Department of Biomedical Engineering,\@ Ben Gurion
  University of the Negev\\ Be\rq er Sheva 84105, Israel}

\author{Oded Farago}

\affiliation{Department of Biomedical Engineering,\@ Ben Gurion
  University of the Negev\\ Be\rq er Sheva 84105, Israel}
\email{ofarago@bgu.ac.il}
%\begin{document}

%\begin{frontmatter}

%\preprint{APS/123-QED}

\begin{abstract}
Membrane proteins often form dimers and higher-order oligomers whose stability and spatial organization depend sensitively on their lipid environment. To investigate the physical principles underlying this coupling, we employ a lattice Monte Carlo model of ternary lipid mixtures that exhibit liquid-disordered ($L_d$) and liquid-ordered ($L_o$) phase coexistence. In this framework, proteins are represented as small membrane inclusions with tunable nearest-neighbor interactions with both lipids and other proteins, allowing us to examine how protein–lipid affinity competes with protein–protein interactions and lipid–lipid demixing. We find that the balance of these interactions controls whether proteins remain dispersed, assemble into small oligomers, or form large stable clusters within $L_o$ domains, and that increasing the protein concentration further promotes coarsening of the ordered phase. To incorporate ligand-regulated activation, we extend the model to a kinetic Monte Carlo scheme in which proteins stochastically switch between inactive and active states with distinct affinities. The inverse switching rate, relative to the time required for a protein to diffuse across the characteristic size of the $L_o$ domains, governs the aggregation behavior. Rapid switching yields only transient small oligomers, slow switching reproduces the static limit with persistent large clusters, and intermediate rates produce broad cluster-size distributions. These results highlight the interplay between lipid phase organization, protein–lipid affinity, and activation dynamics in regulating membrane protein oligomerization, a coupling that is central to signal transduction and membrane organization in living cells.
\end{abstract}
\maketitle
%\begin{sigstatement}
%Membrane protein clustering plays key roles in signaling and regulation, yet the physical factors governing whether proteins remain monomeric or assemble into oligomers are not fully understood. Using a lattice-based model of ternary lipid mixtures with liquid-ordered/liquid-disordered coexistence, we show how protein–lipid affinity, protein–protein attraction, and activation-state dynamics together determine cluster size and stability. A key control parameter is the time a protein remains in its binding-competent state relative to the time required to diffuse across an ordered domain. This work provides a statistical-mechanical framework for understanding how cells may modulate protein oligomerization through changes in local lipid composition, ligand binding, or biochemical regulation, linking molecular interactions to mesoscale membrane organization.
%\end{sigstatement}
%\end{frontmatter}
%\maketitle
\section*{Introduction}

Lipid membranes constitute the boundary that protects biological cells. They confer structural integrity to the cell and its organelles, and play crucial roles in many membrane-associated functions. The main structural constituents of membranes are lipids and proteins. Lipids form the bilayer matrix that serves as a dynamic scaffold, providing lateral fluidity, flexibility, and the capacity for curvature and domain formation~\cite{nagle2000structure,hianik1995bilayer}. The chemical diversity of lipid species, varying in headgroups, tail lengths, and degrees of saturation, enables membranes to regulate local environments and modulate protein activity~\cite{shevchenko2010lipidomics,SHEN2014359}. While hundreds of lipid species have been identified in cell membranes, proteins of various shapes and sizes typically occupy about 25–35\% of the membrane surface area~\cite{Alberts2017,guigas2016effects}. Membrane proteins have been extensively investigated owing to their central roles in membrane-associated processes including, selective molecular transport, signal transduction, and cell–cell communication~\cite{Alberts2017,cheng2019biological,konings1996transport}. They also play organizational and mechanical roles, such as mediating adhesion to the extracellular matrix and interactions with neighboring cells~\cite{jelokhani2022membrane,whitford2013proteins}. These proteins can be deeply embedded within the lipid bilayer or loosely associated with its cytoplasmic or extracellular surfaces. The former are termed integral proteins, while the latter are peripheral proteins~\cite{Alberts2017,klein1985detection}. Integral proteins often span the entire membrane one or multiple times, with the hydrophobic portion forming the transmembrane domain~\cite{von2006membrane}. Because of these domains, integral proteins are difficult to remove without disrupting the membrane~\cite{speers2007proteomics}. In contrast, peripheral proteins are attached to the membrane surface primarily through non-covalent interactions and can, therefore, be more easily detached~\cite{klein1985detection}. In addition, there are lipid-anchored proteins, which are covalently attached to lipid moieties, thereby remaining stably associated with the membrane while retaining lateral mobility~\cite{cross1990glycolipid,saha2016gpi}.

The spatial distribution of lipids within membrane leaflets is heterogeneous, with domains of distinct composition coexisting within them. These domains can be categorized into liquid ordered ($L_o$), liquid disordered ($L_d$), and gel ($S_o$) phases~\cite{mouritsen2010liquid}. The lipid raft represents the most widely recognized example of an $L_o$ domain~\cite{schuck2003resistance}. Rafts are typically 10–200 nm in size, enriched in cholesterol and saturated lipids, mostly sphingolipids, and often contain specific proteins~\cite{pike2006rafts}. They are highly dynamic, constantly assembling and disassembling on short timescales, which allows them to serve as transient platforms for protein sorting, signal transduction, and membrane trafficking~\cite{pike2006rafts}. Because of their central role in organizing membrane proteins and mediating cellular communication, lipid rafts are considered key structural and functional units of the plasma membrane. However, the enormous chemical diversity and organizational complexity of biological membranes make them extremely challenging to study theoretically. To better understand the thermodynamic behavior and mechanical properties that govern domain formation and membrane organization, simplified model systems are often employed. In particular, ternary lipid mixtures containing saturated and unsaturated lipids together with cholesterol have been extensively investigated, as they capture essential features of phase separation while reducing the complexity of the native membrane~\cite{marsh2013handbook}. Phase diagrams have been determined for many such mixtures, and it has been found that they often display a broadly similar structure~\cite{feigenson2009phase,komura2014physical,veatch2007critical,hirst2011phase}. A common feature is the coexistence region between $L_d$ and $L_o$ phases, which serves as the model analogue of raft-like domains in biological membranes.

In addition to the lateral arrangement of lipids, the spatial organization of proteins in membranes is also highly complex. The activity and regulation of many membrane proteins depend on their association with other biomolecules. In many cases, these interacting molecules are other proteins, with numerous membrane proteins self-associating or binding to different partners to form dimers and larger oligomers. An often-cited example is the G-protein coupled receptor (GPCR). GPCRs are ubiquitous in eukaryotic cells, and comprise the largest and perhaps most diverse family of transmembrane receptor proteins~\cite{rios2001g,yang2021g}. They respond to a wide variety of signaling molecules and frequently exist as dimers or higher-order oligomers, which are necessary for their proper functioning~\cite{FARONGORECKA2019155,gurevich2008and}. GPCR assemblies can be either homo-oligomeric (composed of identical GPCR molecules) or hetero-oligomeric (involving different GPCR types)~\cite{milligan2003gpcr,casado2009gpcr}. In addition to GPCRs, the urokinase-type plasminogen activator receptor (uPAR) and several immune receptors, such as T-cell receptors, also exhibit clustering behavior~\cite{cunningham2003dimerization,wucherpfennig2010structural}. Membrane proteins also associate with small molecules, termed ligands or agonists, whose binding often initiates protein dimerization and oligomerization. Ligand binding, for instance, has been shown to promote clustering of GPCRs, as in the case of ligand-induced clustering of N-formyl peptide receptors~\cite{rios2001g,xue2004n}. Similarly, ligand binding has been shown to influence uPAR dimerization, with both positive and negative effects reported~\cite{ge2014ligand}. 

One important facet of protein dimerization and oligomerization is that it can alter the association of proteins with lipid rafts. It has been reported that the affinity of many GPCRs for lipid rafts can either increase or decrease following dimerization~\cite{feron1997dynamic,ishizaka1998angiotensin,xue2004n,huang2007cholesterol,rybin2000differential,fallahi2009lipid}. For example, P2Y12, a member of the GPCR family, associates with lipid rafts in its oligomeric form but relocates out of rafts when present as a monomer~\cite{savi2006active}. Similarly, the translocation of N-formyl peptide receptors to lipid rafts after ligand binding has been suggested, whereas the presence of agonists lowers the affinity of delta opioid receptors for rafts~\cite{xue2004n,huang2007cholesterol}. The previously mentioned uPAR proteins also associate with lipid rafts upon dimerization~\cite{cunningham2003dimerization}. The interplay between ligand-induced dimerization and changes in raft affinity is not restricted to GPCRs. T-cell receptors (TCRs), for instance, have little affinity for lipid rafts in their resting state, but after antigen binding they translocate into rafts, where they interact with other proteins for chemical modification~\cite{horejsi2003roles,he2005lipid,kabouridis2006lipid}. Epidermal growth factor receptors (EGFRs), members of the ErbB family, can also exist in monomeric or dimeric forms, with ligand binding enhancing dimer stability~\cite{chung2010spatial,lemmon2010cell,singh2020revisiting}. The relationship between EGFRs and lipid rafts is cell-type dependent, which has led to conflicting findings on their membrane distribution~\cite{ruzzi2024lipid}. In the glioblastoma cell line U87MG, EGFRs colocalize with lipid rafts, and ligand binding disrupts this association~\cite{ruzzi2024lipid,abulrob2004interactions}. In contrast, in the A431 cell line, EGFRs show little raft affinity, with only about 40\% residing in rafts, and ligand binding does not significantly alter this distribution~\cite{ruzzi2024lipid}.

Thus, both protein–protein and protein–ligand interactions are fundamental mechanisms regulating the spatial organization and functional dynamics of biological membranes.
The forces that drive the formation and stabilization of membrane protein complexes are diverse and predominantly noncovalent~\cite{dill1990dominant}. A key contribution arises from the hydrophobic effect. Nonpolar residues of proteins tend to aggregate to minimize their exposure to water, often leading to supramolecular assemblies~\cite{lins1995hydrophobic,southall2002view,spolar1989hydrophobic}. Electrostatic interactions and van der Waals forces also play major roles, while specific docking interfaces are frequently stabilized by hydrogen bonds~\cite{zhou2018electrostatic,roth1996van,gorbenko2011protein}. In addition, $\pi$–$\pi$ stacking between aromatic residues and metal coordination can provide further stabilization~\cite{mizanur2015aromatic,tao2023interface}. Entropic effects also contribute, including the release of ordered water molecules and counterions upon complex formation, as well as membrane curvature–mediated interactions, which usually favor clustering~\cite{banerjee2020dynamical,thirumalai2012role,bruinsma1996protein,farago1,farago2,harries2013counterion}. Lipids can modulate protein conformation and orientation, and generate local stresses that influence protein–protein interactions, and act as mediators by forming annular shells or lipid bridges between proteins~\cite{kurouski2023elucidating,muller2019characterization,de2008molecular,marvcelja1976lipid}.

A limited number of \textit{in silico} studies have examined membrane protein dimerization. Kargar \textit{et al.} investigated the interactions of two $\beta$-amyloid peptides in aqueous solution and within a DPPC bilayer, showing that dimerization perturbs lipid packing and disrupts bilayer order~\cite{kargar2020dimerization}. Multiscale simulations combining MARTINI coarse-grained and atomistic models have been used to study how the transmembrane helices of synaptobrevin-2 and of the thyrotropin receptor associate with identical partners to form homodimers in a lipid bilayer~\cite{han2015synaptobrevin,mezei2022modeling}. The free energy of dimerization of the NanC membrane protein was also estimated in a phospholipid bilayer using umbrella sampling within the MARTINI framework~\cite{dunton2014free}. However, the tendency of this model to over-aggregate membrane proteins has been noted, leading to several proposed reparameterizations of the force field~\cite{majumder2021addressing}. These examples highlight the complementary strengths of atomistic and coarse-grained simulations: the former provide molecular detail but are limited in time and length scales, whereas the latter enable broader sampling of larger systems. Nonetheless, even coarse-grained models such as MARTINI remain restricted in accessible size and duration compared with the biological processes they aim to capture.

To overcome these limitations, ultra–coarse-grained (UCG) methods have been employed to investigate protein–protein and protein–lipid interactions~\cite{brannigan2007contributions,hu2012determining,west2009membrane,argudo2016continuum,illya2008coarse,argudo2017new,kahraman2016bilayer,gao2021membrane,morriss2014coarse,lee2018kinetic}. Lattice models of membranes with proteins offer an even higher level of abstraction, reducing the molecular structures of lipids and proteins to effective, typically nearest-neighbor, interactions between lattice sites representing different molecules. This simplification enables simulations on scales that are macroscopically large, and entirely beyond the reach of molecular UCG simulations. Early lattice-based Monte Carlo modeling efforts explored GPCR dimerization and the influence of lipid rafts, with proteins represented as hexagons and rafts treated as low-diffusivity lattice regions lacking explicit lipid detail~\cite{woolf2003self,fallahi2009lipid}. More recently, we developed a lattice model of ternary lipid mixtures composed of saturated and unsaturated lipids together with cholesterol, in which lipids are represented as dimers to reflect their two hydrocarbon tails, while cholesterol is represented as monomers~\cite{sarkar_prr,sarkar_soft,sarkar2023}. The model successfully reproduces the phase behavior of ternary lipid mixtures, capturing both macroscopic separation characterized by large liquid-ordered domains and microscopic separation involving many small liquid-ordered domains.Building on this framework, we recently examined the effect of incorporating small proteins (peptides) into ternary mixtures and found that protein aggregation strongly influences the size and morphology of the liquid-ordered domains. In the same study, we observed that in mixtures already exhibiting $L_d$+$L_o$ coexistence, proteins partition differently between the two phases, where aggregation promotes lipid recruitment and phase separation~\cite{basujcp}.

In the present work, we extend our previous lattice model of ternary lipid mixtures with small proteins to investigate the formation of different types of protein clusters, ranging from dimers to oligomers and larger assemblies. To this end, we introduce an explicit protein–protein interaction term, which allows us to explore how varying interaction strength affects protein cluster formation. We also employ kinetic Monte Carlo simulations to account for the dynamic nature of binding–activation events. In this framework, proteins switch between ‘active’ and ‘passive’ states, where activation represents binding to small molecules or regulatory proteins, leading to stronger protein–protein interactions within the membrane.  Finally, we also allow the affinity of proteins for $L_o$ domains to depend on the activation state, thereby linking clustering behavior to membrane lateral organization. This approach aims to capture, within a simple physical model, essential aspects of how dynamic binding and interaction mechanisms shape membrane protein organization.

\section*{Methods}

The details of the Monte Carlo (MC) scheme are provided in ref.~\cite{basujcp}. 
Here, we briefly recapitulate the lattice model, in which ternary mixtures of saturated 
and unsaturated lipids with cholesterol (Chol) are mapped onto a triangular lattice 
of $N_0=121 \times 140 = 16\,940$ sites with periodic boundary conditions. Lipids are modeled 
as dimers to reflect their two hydrocarbon tails, cholesterol as monomers, and a fraction 
of sites are left empty (‘voids’) to adjust average density differences between phases; 
the lattice spacing corresponds to $l\simeq 0.56$~nm. In addition, small proteins (‘peptides’) 
are included as triangle-shaped trimers, with each 100 peptides corresponding to an area 
coverage of approximately 2\% of the lattice.

\begin{figure*}
    \includegraphics[width=0.8875\textwidth]{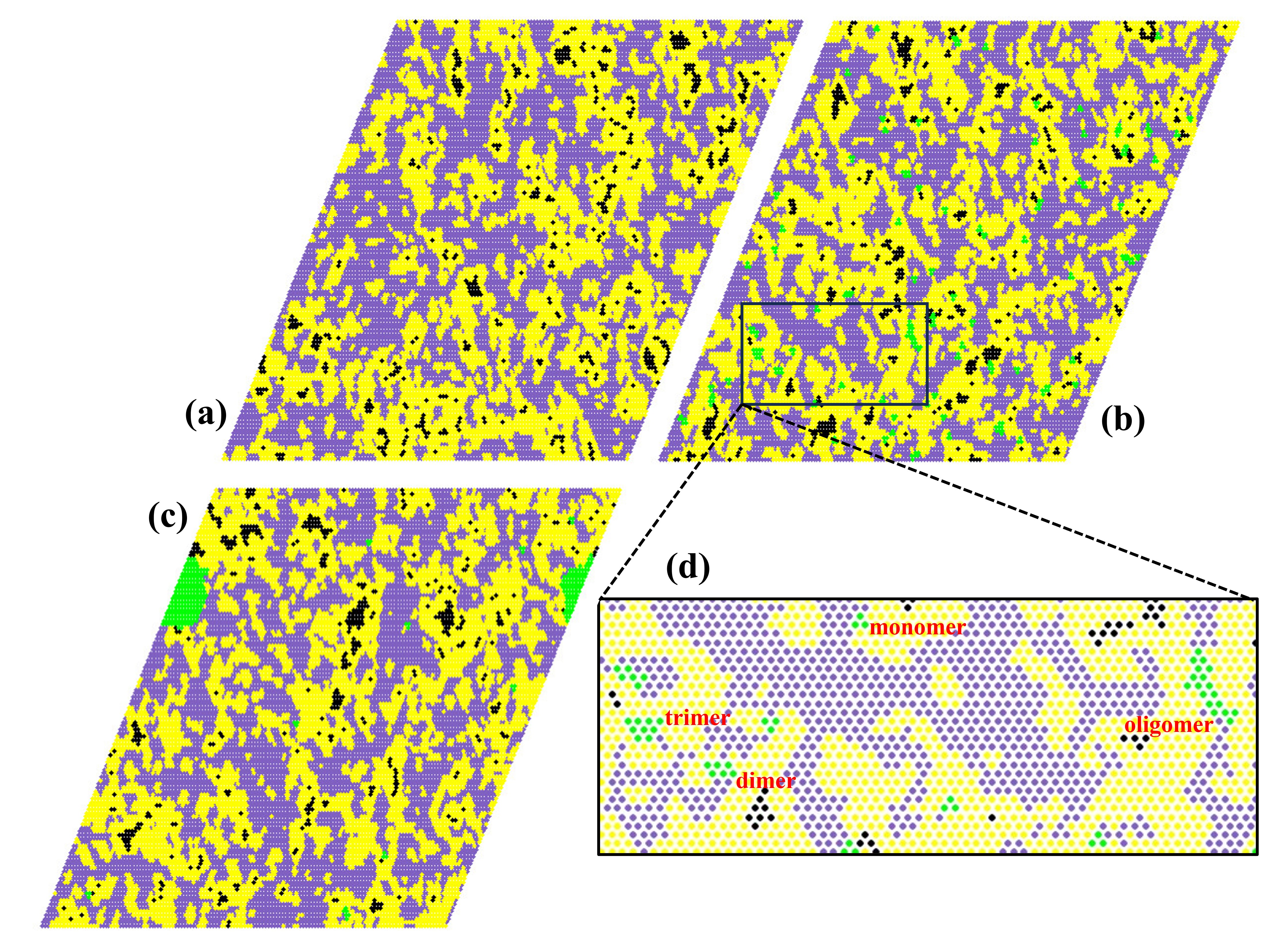}
    \centering
    \caption{Equilibrium snapshots of the protein–lipid system (35DPPC): (a) no proteins, (b) $N=100$ proteins with interactions $\varepsilon_{25}=0.75$ and $\varepsilon_{55}=0.75$, (c) $N=100$ proteins with $\varepsilon_{25}=0.75$ and $\varepsilon_{55}=1.3$. (d) An enlarged view of the marked section of panel (b). Liquid-ordered ($L_o$), liquid-disordered ($L_d$), gel ($S_o$), and proteins are shown in yellow, purple, black, and green, respectively.}
    \label{pic:prosnap}
\end{figure*}

The 6-state model assigns to each lattice site one of the following states: 
$s=0$ (void), $s=1$ (disordered saturared chain), $s=2$ (ordered saturated chain), 
$s=3$ (Chol), $s=4$ (unsaturated chain), and $s=5$ (protein/peptide).
The model Hamiltonian reads
\begin{eqnarray}
E &=& -\Omega k_B T \sum_i \delta_{s_i,1} 
      - \sum_{\langle i,j \rangle} \Big[
      \varepsilon_{22}\,\delta_{s_i,2}\delta_{s_j,2} 
      + \varepsilon_{23}\,\delta_{s_i,2}\delta_{s_j,3} \nonumber \\[6pt]
  &+& \varepsilon_{24}\,\delta_{s_i,2}\delta_{s_j,4} 
      + \varepsilon_{25}\,\delta_{s_i,2}\delta_{s_j,5}
      + \varepsilon_{55}\,\delta_{s_i,5}\delta_{s_j,5} \Big],
      \label{eq:hamiltonian}
\end{eqnarray}
where $\langle i,j\rangle$ denotes nearest–neighbor pairs and $\varepsilon_{ab}=\varepsilon_{ba}$. 
We set $\Omega=3.9$, $\varepsilon_{22}=1.3\varepsilon$, $\varepsilon_{23}=0.72\varepsilon$, and $\varepsilon_{24}=0.40\varepsilon$, where the energy unit is defined as $\varepsilon = k_B T_m / 0.9$, with $T_m \simeq 314~\mathrm{K}$ being the melting temperature of a DPPC membrane. All energies in this work are expressed in units of $\varepsilon$.
At $T\simeq 300~\mathrm{K}$, this parameter set reproduces the phase diagram of the ternary DPPC (saturated)/DOPC (unsaturated)/Chol mixture, showing $L_d+L_o$ coexistence with liquid-ordered domains of several to a few tens of nanometers in size that are often interconnected [see simulation snapshot in Fig.~\ref{pic:prosnap}(a)].

%We set $\Omega=3.9$, $\varepsilon_{22}=1.3\varepsilon$, $\varepsilon_{23}=0.72\varepsilon$, and $\varepsilon_{24}=0.40\varepsilon$. At $T\simeq 300~\mathrm{K}$, this parameter set reproduces the phase diagram of the ternary DPPC (saturated)/DOPC (unsaturated)/Chol mixture, showing $L_d+L_o$ coexistence with liquid-ordered domains of several to a few tens of nanometers in size that are often interconnected [see simulation snapshot in Fig.~\ref{pic:prosnap}(a)]. We use energy units with $\varepsilon = k_B T_m/0.9$, where $T_m \simeq 314~\mathrm{K}$  is the melting temperature of a DPPC membrane, and all energies in the following are  expressed in units of $\varepsilon$.

In addition, we run kinetic Monte Carlo (KMC) simulations aimed at investigating the influence of ligand-binding activation of the proteins. In this set of simulations, activation modifies the proteins’ self-interactions as well as their affinities toward the different lipid species. In the kinetic simulations, a protein can switch between state $s=5$, referred to as the unbound state, and an “active” bound state $s=6$. 
The active state is characterized by modified interaction parameters, specifically $\varepsilon_{26}$ for interactions with ordered DPPC chains, and $\varepsilon_{66}$ for protein–protein interactions. Accordingly, we add in these simulations the term $ \varepsilon_{26}\,\delta_{s_i,2}\delta_{s_j,6}+ \varepsilon_{66}\,\delta_{s_i,6}\delta_{s_j,6}$ to the Hamiltonian (\ref{eq:hamiltonian}). To keep minimal the number of control parameters studied, we assume no direct interactions between the proteins in these two states, i.e., we set $\varepsilon_{56}=0$. The proteins switch between the two states at constant activation $k_{\rm on}$,  and deactivation $k_{\rm off}$, rates. This is implemented by drawing the time intervals between subsequent switches from exponential distributions: from the inactive state $s=5$ according to $p(t)=k_{\rm on}\exp(-k_{\rm on}t)$, and from the active state $s=6$ according to $p(t)=k_{\rm off}\exp(-k_{\rm off}t)$, where $t$ is measured in Monte Carlo time units. In steady state, the proteins partition between states $s=5$ and $s=6$ with a ratio $r=k_{\rm off}/k_{\rm on}$. Nevertheless, the dynamics and the size distribution of the protein assemblies (i.e., the number of proteins involved in an assembly) depend on the individual values of $k_{\rm off}$ and $k_{\rm on}$, not only on their ratio. In the kinetic simulations, we vary the activation and deactivation rates while keeping all other model parameters fixed. As a reference, we define $k_0 = 1/\tau_0 = 1/30000\;(\text{MC time})^{-1}$, where $\tau_0$ is the measured average time for a single protein to diffuse across one quarter of the lattice (approximately 30 lattice spacings in each direction; see Fig.~S1 in the Supporting Information). This distance corresponds to roughly 16 nm, which is comparable to the typical size of the $L_o$ domains and allows a protein to encounter several others during this time. Using $k_0$ as a reference rate provides a clear distinction between slow ($k \ll k_0$) and fast ($k \gg k_0$) switching regimes. For the record, one Monte Carlo (MC) time unit in our simulations corresponds to a sequence of $N_0=16,940$ attempted moves (equal to the number of lattice sites), of which on average $95\%$ are molecular displacements and $5\%$ are state changes of ordered lipid chains ($s=1 \leftrightarrow 2$). A typical run spans $2.4\times 10^{7}$ MC time units and requires approximately 8 days of CPU time.

We present numerous simulation snapshots in which the different phases are color-coded as in our previous studies: purple for $L_d$, yellow for $L_o$, and black for $S_o$. The algorithm used to classify lattice sites by phase is described in ref.~\cite{basujcp}. Proteins are shown in green, and any additional color coding is specified in the respective figure captions.

\section*{Results and Discussion}

%Both static and kinetic model of interacting proteins in lipid mixture has been penned in this section.  

\subsection*{MC simulations of interacting proteins}

We begin with MC simulations of mixtures containing molecular fractions of 
35\%:40\%:25\% DPPC/DOPC/Chol (hereafter referred to as the 35DPPC mixture). 
In the absence of proteins, and with the model parameters used, the 35DPPC mixture exhibits $L_d$+$L_o$ phase coexistence, where liquid-ordered domains form a loosely connected network embedded in the liquid-disordered matrix, with typical domain sizes of about 10 nm [Fig.~\ref{pic:prosnap}(a)]. We simulate this lipidic mixture with initially $N=100$ proteins  added, focusing on the formation and size distribution of protein assemblies within the $L_o$ phase. These are governed by the interplay of diverse short-range interactions between
the saturated lipids ($s=2$) and proteins ($s=5$), which are represented in our model by the parameters $\varepsilon_{22}=1.3$, $\varepsilon_{25}$, and $\varepsilon_{55}$.
Figures~\ref{pic:prosnap}(b) and (c) show snapshots of the mixture containing 100 proteins, which display protein aggregation within liquid-ordered domains but of markedly different organization. In (b), we observe many small aggregates, mostly dimers and trimers, along with a substantial fraction of unbound monomers, whereas in (c) the proteins form a large cluster that includes almost all the proteins present.

\subsubsection*{Size distribution of protein clusters}

\begin{figure*}
    \includegraphics[width=0.8875\textwidth]{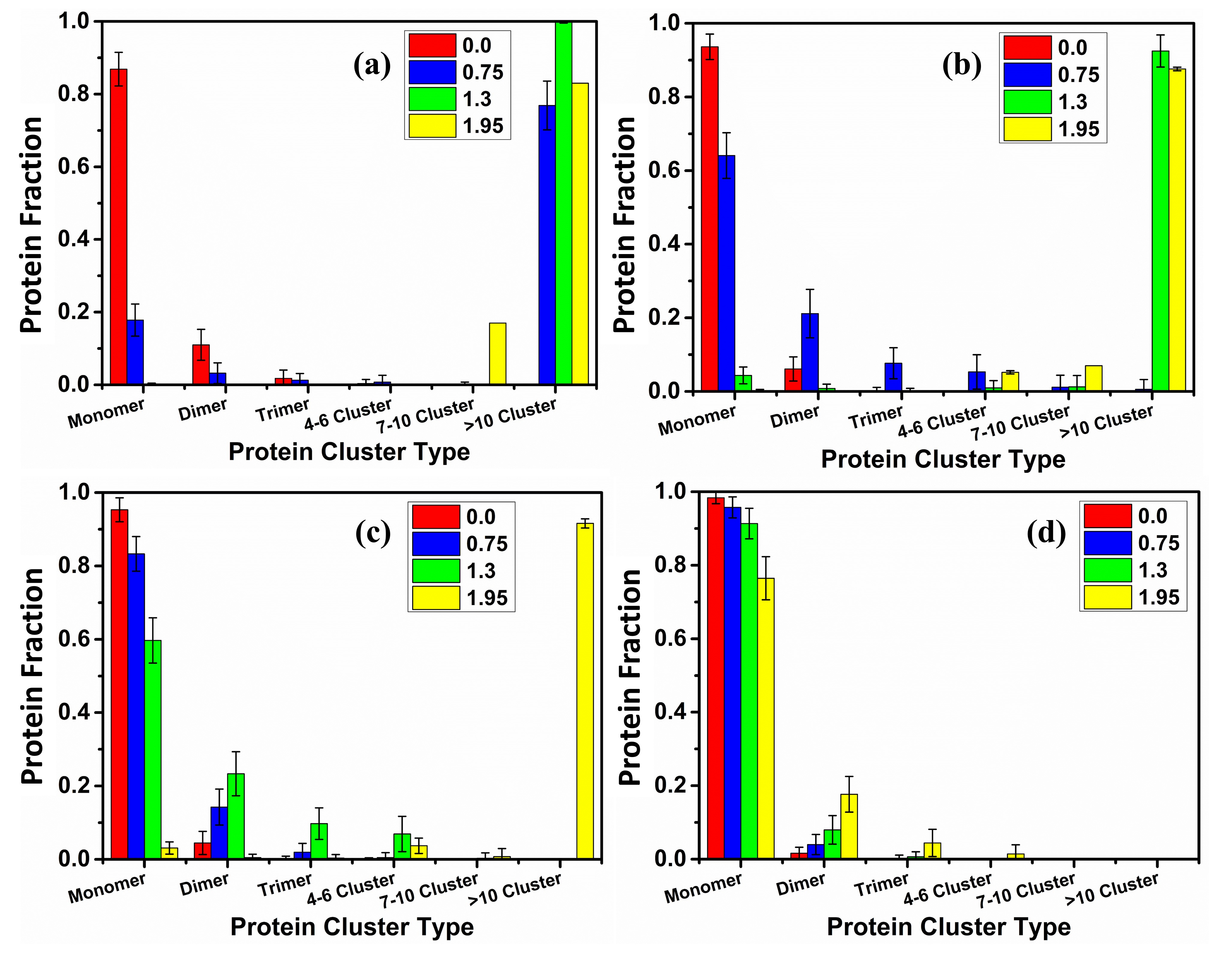}
    \centering
    \caption{Fraction of proteins belonging to different cluster sizes for various values of $\varepsilon_{55}$ (indicated by different colors in the insets) and (a) $\varepsilon_{25}=0$, (b) $\varepsilon_{25}=0.75$, (c) $\varepsilon_{25}=1.3$, and (d) $\varepsilon_{25}=1.95$. The categories ‘4–6 cluster,’ ‘7–10 cluster,’ and ‘$>10$ cluster’ denote aggregates containing 4–6, 7–10, and more than 10 proteins, respectively. Error bars represent standard deviations obtained from independent simulation snapshots.}
    \label{pic:procluster}
\end{figure*}

Quantitative data on the size distribution of protein aggregates are provided in Fig.~\ref{pic:procluster}, which shows histograms of the fraction of proteins belonging to different cluster sizes (monomers, dimers, oligomers, etc.) for various values of $\varepsilon_{25}$ and $\varepsilon_{55}$. As expected, in the absence of protein–protein attraction ($\varepsilon_{55}=0$), proteins are predominantly found in the unbound monomeric state, with only a small fraction forming dimers and no larger clusters observed. This behavior is consistent across all values of $\varepsilon_{25}$ investigated. For small $\varepsilon_{55}$ values, or when $\varepsilon_{55}$ is close to $\varepsilon_{25}$, a significant fraction of proteins form small clusters (dimers and trimers), while larger clusters remain negligible. For example, when $\varepsilon_{25}=\varepsilon_{55}=0.75$ [fig.~\ref{pic:procluster}(b)], about 40\% of proteins participate in small clusters, of which roughly 20\% are dimers. Nearly identical fractions of monomers, dimers, trimers, and small (4–6) oligomers are observed for $\varepsilon_{25}=\varepsilon_{55}=1.3$ [fig.~\ref{pic:procluster}(c)], and similar distributions of cluster sizes are also found for $\varepsilon_{25}=\varepsilon_{55}=1.95$ [fig.~\ref{pic:procluster}(d)]. 
A qualitatively different behavior emerges when $\varepsilon_{55}$ significantly exceeds $\varepsilon_{25}$. In this case, large aggregates ($>10$ proteins) become the dominant cluster type, while the fractions of monomers and small clusters decrease sharply. For instance, for $\varepsilon_{25}=1.3$ and $\varepsilon_{55}=1.95$ [fig.~\ref{pic:procluster}(c)], more than 80\% of proteins belong to large clusters, while the remaining fraction is distributed among monomers and smaller oligomers. This trend highlights the cooperative nature of protein clustering once the protein–protein attraction becomes stronger than the lipid–protein coupling.
%At sufficiently high $\varepsilon_{55}$, large clusters ($>10$ proteins per cluster) become the  dominant form, with only a minimal presence of smaller clusters. An example is  the system with $\varepsilon_{25}=1.3\varepsilon$ and $\varepsilon_{55}=1.95\varepsilon$.

These results delineate three distinct regimes of protein organization within the lipid mixture. At low protein–protein attraction ($\varepsilon_{55}<\varepsilon_{25}$), proteins remain dispersed, with clustering energetically unfavorable. When $\varepsilon_{55}\approx\varepsilon_{25}$, small aggregates such as dimers and trimers become favorable, while larger clusters are still suppressed. Finally, when $\varepsilon_{55}>\varepsilon_{25}$, large protein clusters dominate the system and other forms become negligible. This energetic balance is also strongly coupled to the diffusive dynamics: weak protein–protein interactions allow relatively unhindered diffusion, leading primarily to small clusters formed through transient encounters, whereas strong interactions hinder diffusion and slow the dynamics but, at the same time, promote the merging of proteins into large, stable clusters. This classification underscores the central role of the competition between protein–lipid ($\varepsilon_{25}$) and protein–protein ($\varepsilon_{55}$) interactions in regulating protein self-assembly within membranes.

\subsubsection*{Partitioning of proteins between phases}

\begin{figure*}
    \includegraphics[width=0.8875\textwidth]{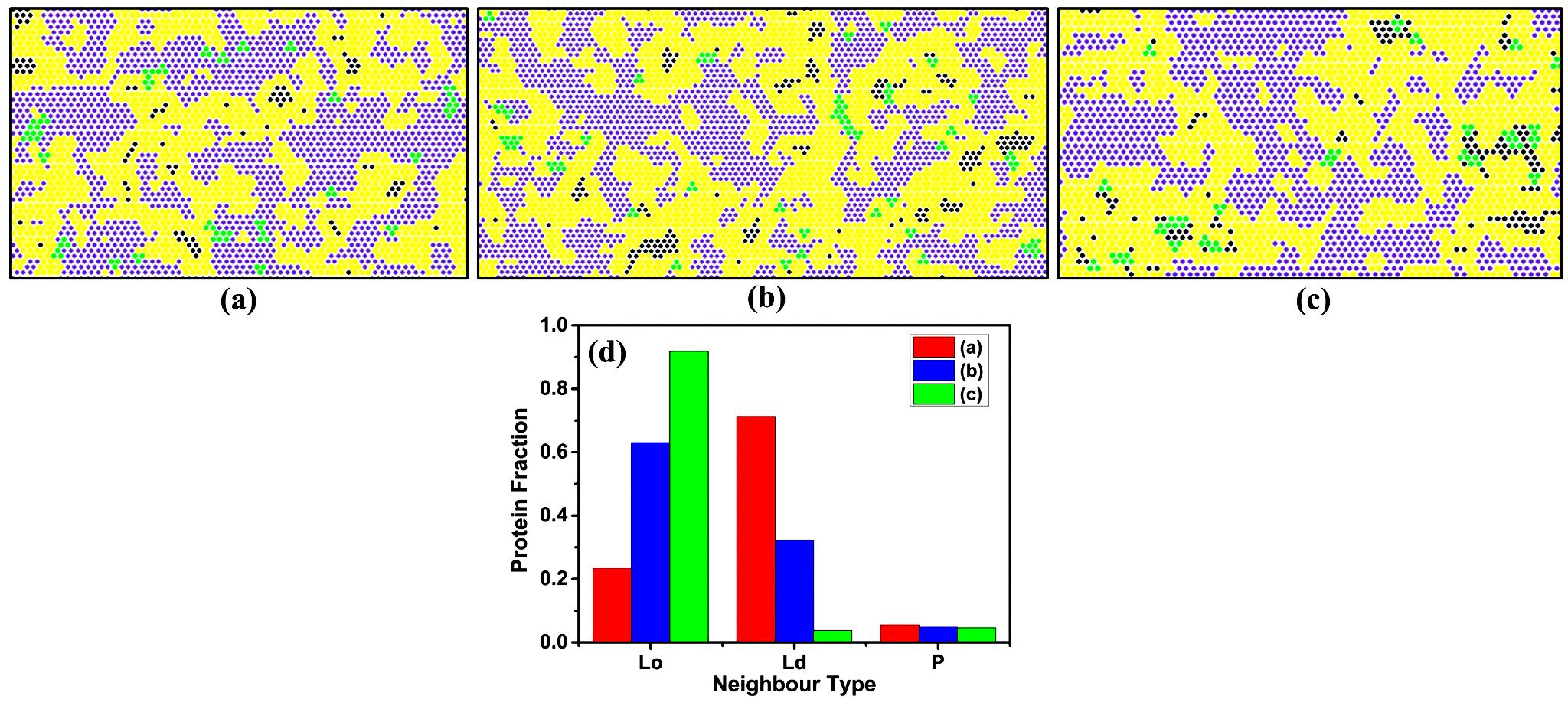}
    \centering
    \caption{ Zoomed-in regions of 35DPPC mixtures showing the locations of small proteins for (a) $\varepsilon_{25}=0$, $\varepsilon_{55}=0.5$, (b) $\varepsilon_{25}=\varepsilon_{55}=0.75$, and (c) $\varepsilon_{25}=\varepsilon_{55}=1.3$. The color coding is the same as in Fig.~\ref{pic:prosnap}. (d) Corresponding histograms of the distribution of protein neighbors among the different components ($L_o$, $L_d$, and proteins - P) for the systems shown in (a–c).}
    \label{pic:neib_stat}
\end{figure*}

Apart from their size, the locations of protein clusters are also important, since protein functionality depends on the surrounding lipid environment. While some proteins preferentially localize in the $L_o$ phase, others are more often found in the $L_d$ phase. Consistent with our previous simulations of systems with non-interacting proteins ($\varepsilon_{55}=0$)~\cite{basujcp}, we find that the partitioning of proteins between the two phases is mainly governed by the protein–DPPC affinity parameter, $\varepsilon_{25}$. Figure~\ref{pic:neib_stat} illustrates this dependence. The snapshots show equilibrium configurations of systems with (a) $\varepsilon_{25}=0$, $\varepsilon_{55}=0.5$, (b) $\varepsilon_{25}=\varepsilon_{55}=0.75$, and (c) $\varepsilon_{25}=\varepsilon_{55}=1.3$. These three systems exhibit relatively similar populations of protein clusters consisting mainly of monomers and dimers with a few oligomers, which is characteristic of mixtures with $\varepsilon_{55}\approx\varepsilon_{25}$. The snapshots highlight a gradual shift in the locations of the small clusters: for small $\varepsilon_{25}$ values (a), they are primarily found in the $L_d$ domains, with some at the extended $L_o/L_d$ interfaces. This reflects the fact that when $\varepsilon_{25}<\varepsilon_{22}=1.3$, the stronger DPPC–DPPC attraction in the $L_o$ phase reduces the likelihood of protein localization there. At $\varepsilon_{25}=0.75$ (b), proteins are distributed more evenly between the $L_o$ and $L_d$ phases, with many localized along the phase boundaries. For $\varepsilon_{25}=1.3$ (c), where protein–DPPC affinity matches the DPPC–DPPC interaction strength, most proteins localize inside the $L_o$ domains. This shift in localization as $\varepsilon_{25}$ increases is the same mechanism identified previously~\cite{basujcp}, where the protein–DPPC affinity was shown to be the primary determinant of phase preference.
Panel (d) quantifies these trends by showing the fractions of protein neighbors found within $L_o$, $L_d$, and protein (P) environments. The data illustrate the gradual transfer of proteins from $L_d$ to $L_o$ with increasing $\varepsilon_{25}$. They also show a small fraction of protein neighbors that remains nearly identical in all three cases, reflecting the broadly similar cluster populations.

\subsubsection*{Influences of protein density}
\begin{figure*}
    \includegraphics[width=0.8875\textwidth]{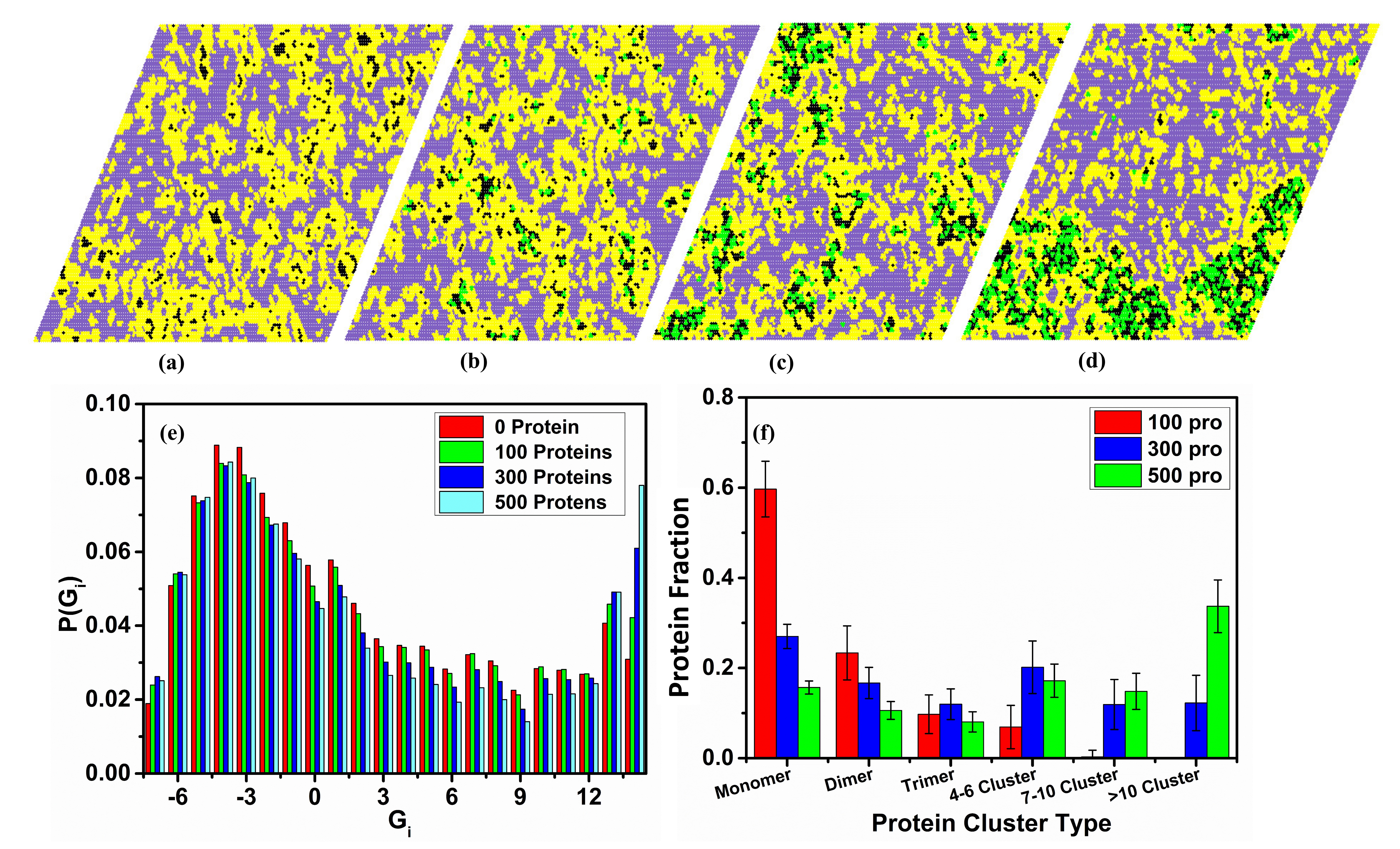}
    \centering
    \caption{(a–d) Equilibrium snapshots of 35DPPC lipid mixtures with $\varepsilon_{25}=\varepsilon_{55}=1.3$ containing (a) $N=0$, (b) $N=100$, (c) $N=300$, and (d) $N=500$ proteins. The color coding is the same as in Fig.~\ref{pic:prosnap}. (e) Distribution of the local order parameter $G_i$ for different protein concentrations. (f) Histogram of the fraction of proteins belonging to clusters of different sizes (using the same grouping as in Fig.~\ref{pic:procluster}).}
    \label{pic:pronumvar}
\end{figure*}
%In order to investigate the effects of number of interacting proteins on the protein clustering behaviour and also on the ternary type-I lipid mixture, we have added 300 and 500 proteins in the type-I 35DPPC lipid mixture. Fig.\ref{pic:pro_num} represents the change in the clustering behaviour and neighbour statistics. From Fig.\ref{pic:neib_stat}(c) and Fig.\ref{pic:pro_num}(a)-(b), it is evident that the overall clustering behaviour remains same with the increase of the protein number. The only notable difference is that with the increment of proteins, the larger clusters started to appear at a lower value of $\varepsilon_{55}$. For example, for $\varepsilon_{55}=1.0\varepsilon$, there is no larger protein cluster (7-10/$>$10 clusters) for 100 proteins (Fig.\ref{pic:neib_stat}(c)). For 300 proteins, unstable 7-10 protein clusters were observed in the system  but not $>$10 protein aggregates (Fig.\ref{pic:pro_num}(a). In the case of 500 proteins, both 7-10 and $>$10 protein clusters appeared (Fig.\ref{pic:pro_num}(b)). The underlying reason behind this behaviour is that with a higher number of proteins, it is relatively easy to accumulate in larger clusters even at lower protein-protein interaction energy ($\varepsilon_{55}$). On the other hand, the location of the protein clusters does not change with protein number variation, as seen from Fig.\ref{pic:neib_stat}(c), and Fig.\ref{pic:pro_num}(c). 

To examine how protein density influences clustering and the organization of the 35DPPC lipid mixture, we performed simulations with $\varepsilon_{25}=\varepsilon_{55}=1.3$ and with different numbers of proteins, where each 100 proteins correspond to an area coverage of about 2\%. Figure~\ref{pic:pronumvar}(a)–(d) shows that increasing the protein population progressively drives the nanoscopically phase-separated system toward macroscopic phase separation. In particular, when the system contains 500 proteins, a large macroscopic $L_o$ domain emerges [Fig.~\ref{pic:pronumvar}(d)], which incorporates the proteins and recruits a significant fraction of the saturated lipids. This behavior closely resembles our previous study of systems with non-interacting proteins~\cite{basujcp}. 
In both cases, the dominant driving force is the protein–DPPC affinity $\varepsilon_{25}$, which promotes protein localization within $L_o$ regions, whereas the direct protein–protein interaction $\varepsilon_{55}$ plays only a secondary role.

Panel (e) shows the distribution of the local order parameter $G_i$, which provides a quantitative measure for distinguishing the different lipid environments. Following the classification scheme introduced in ref.~\cite{sarkar_soft} and expanded in~\cite{basujcp}, negative values of $G_i$ correspond to the $L_d$ phase, while lattice sites where $G_i \geq 0$ are associated with the $L_o$ phase. This parameter varies between $-7$ and $14$, with the maximum value $G_i=14$ representing gel-like $S_o$ regions within the $L_o$ domains ("domains within domains"). The existence of these $S_o$-like regions which, in the snapshots, appear as black patches embedded in the yellow liquid-ordered areas, has been established in previous studies~\cite{sarkar2023,sarkar_soft} as a hallmark of the $L_o$ phase in the coexistence regime. This feature distinguishes the $L_o$ phase from the nanoscopic liquid-ordered domains observed in the one-phase regime at lower DPPC content, where such internal gel-like clusters are absent and the domains are simply transient thermal density fluctuations.
As seen in panel (e), the addition of proteins leads to a noticeable increase in the fraction of gel-like regions, reflected by the enhanced weight of the histogram at higher $G_i$ values, and particularly at $G_i=14$. The clustering properties of the proteins also change markedly with the number of proteins, as illustrated in fig.~\ref{pic:pronumvar}(f). At low protein content ($N=100$), proteins are mainly found as monomers, dimers, and small oligomers, while larger assemblies are essentially absent. With increasing $N$, the cluster-size distribution shifts toward larger aggregates, and in the case of $N=500$ nearly one-third of all proteins belong to clusters containing more than 10 proteins. These results demonstrate that the protein density strongly influences both the extent of protein aggregation and the accompanying reorganization of the lipid environment.

\subsection*{Kinetic Model of Interacting Proteins}
\begin{figure*}
    \includegraphics[width=0.9\textwidth]{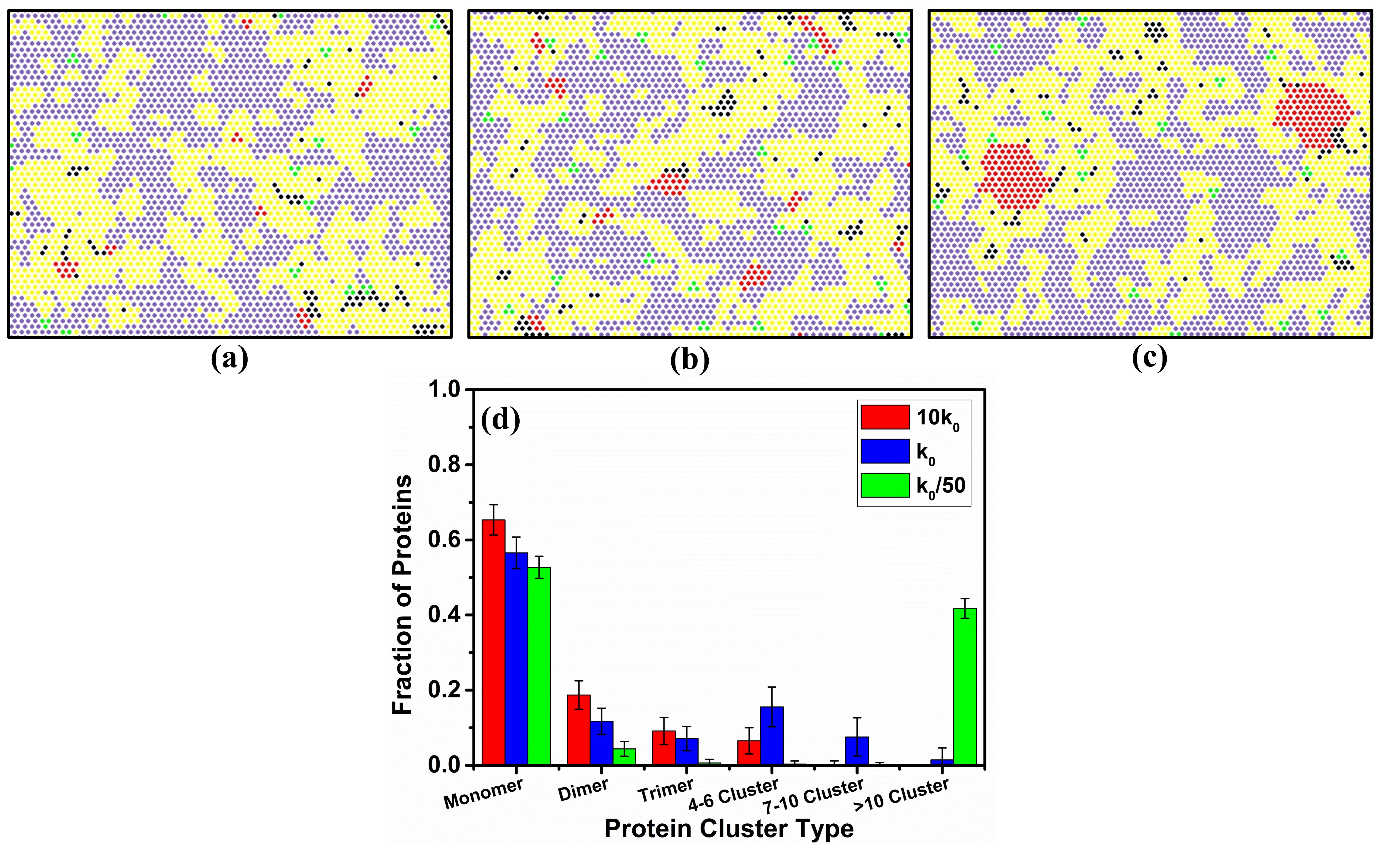}
    \centering
    \caption{Enlarged views of 35DPPC lipid mixtures with $N=200$ proteins, showing the spatial localization of active (red) and inactive (green) proteins for $\varepsilon_{25}=\varepsilon_{26}=0.75$, $\varepsilon_{55}=0$, and $\varepsilon_{66}=1.3$. Snapshots (a), (b), and (c) are obtained for on/off switching rates $k=10k_0$, $k=k_0$, and $k=k_0/50$, respectively. (d) Protein cluster size distributions for the same systems, with error bars representing standard deviations from independent simulation snapshots.}
    \label{pic:kmcsnap}
\end{figure*}

Up to this point, we have considered proteins with fixed states and interactions, unaffected by their environment. In biological membranes, however, proteins can dynamically change their conformations and affinities upon binding or unbinding of specific ligands or agonists. The associated activation and deactivation events influence the statistics of protein aggregate sizes, which may therefore differ from those in the static (fixed-state) case considered earlier. In the kinetic MC simulations, proteins switch stochastically between an inactive state ($s=5$) and an active state ($s=6$), with modified lipid–protein and protein–protein interactions in the active state. The switching dynamics are controlled by fixed activation ($k_{\rm on}$) and deactivation ($k_{\rm off}$) rates, while the residence times in each state are drawn from exponential distributions (see Methods).  The steady-state ratio of inactive to active proteins is $r = k_{\rm on}/k_{\rm off}$. In this work, we set $r = 1$, such that $k_{\rm off} = k_{\rm on} = k$, and vary the rates relative to $k_0 = \tau_0^{-1}$, where $\tau_0$ is the diffusion time of a single protein over one quarter of the lattice (approximately 16 nm).

As a case study for the influence of switching rates on aggregation statistics, we consider $\varepsilon_{25}=\varepsilon_{26}=0.75$, $\varepsilon_{55}=0$, and $\varepsilon_{66}=1.3$. In the static limit, inactive proteins that lack mutual attraction are dominated by monomers, whereas active proteins that interact strongly form large clusters [see Fig.~\ref{pic:procluster}(b)]. We simulate systems with $N=200$ proteins, so that at any given time approximately half ($N/2\approx100$) are active and half inactive. Figure~\ref{pic:kmcsnap} shows equilibrium snapshots for three switching rates: (a) $k=10k_0$, (b) $k=k_0$, and (c) $k=k_0/50$. For the largest switching rate [Fig.~\ref{pic:kmcsnap}(a)], active proteins form only small aggregates, predominantly dimers and trimers. At the intermediate rate [Fig.~\ref{pic:kmcsnap}(b)], moderately sized clusters appear without forming a single dominant aggregate. In all cases, clustering is restricted to the \textit{active} proteins, whereas the \textit{inactive} proteins remain dispersed as monomers. At the smallest switching rate [Fig.~\ref{pic:kmcsnap}(c)], proteins have sufficient time to diffuse across the lattice before changing state, leading to the static-limit configuration in which inactive proteins remain mostly monomeric while active proteins assemble into a large cluster. This behavior is consistent with the static two-state distributions shown in Fig.~\ref{pic:procluster}(b), where inactive proteins ($\varepsilon_{55}=0$) form a monomer-rich population, whereas active proteins ($\varepsilon_{55}=1.3$) form large aggregates. The corresponding cluster-size distributions in Fig.~\ref{pic:kmcsnap}(d) mirror these trends: for $k=k_0/50$, the distribution matches the static-limit behavior; for $k=k_0$, an intermediate broad distribution is observed; and for $k=10k_0$, clusters containing more than six proteins are nearly absent.

\begin{figure*}
   \includegraphics[width=0.9\textwidth]{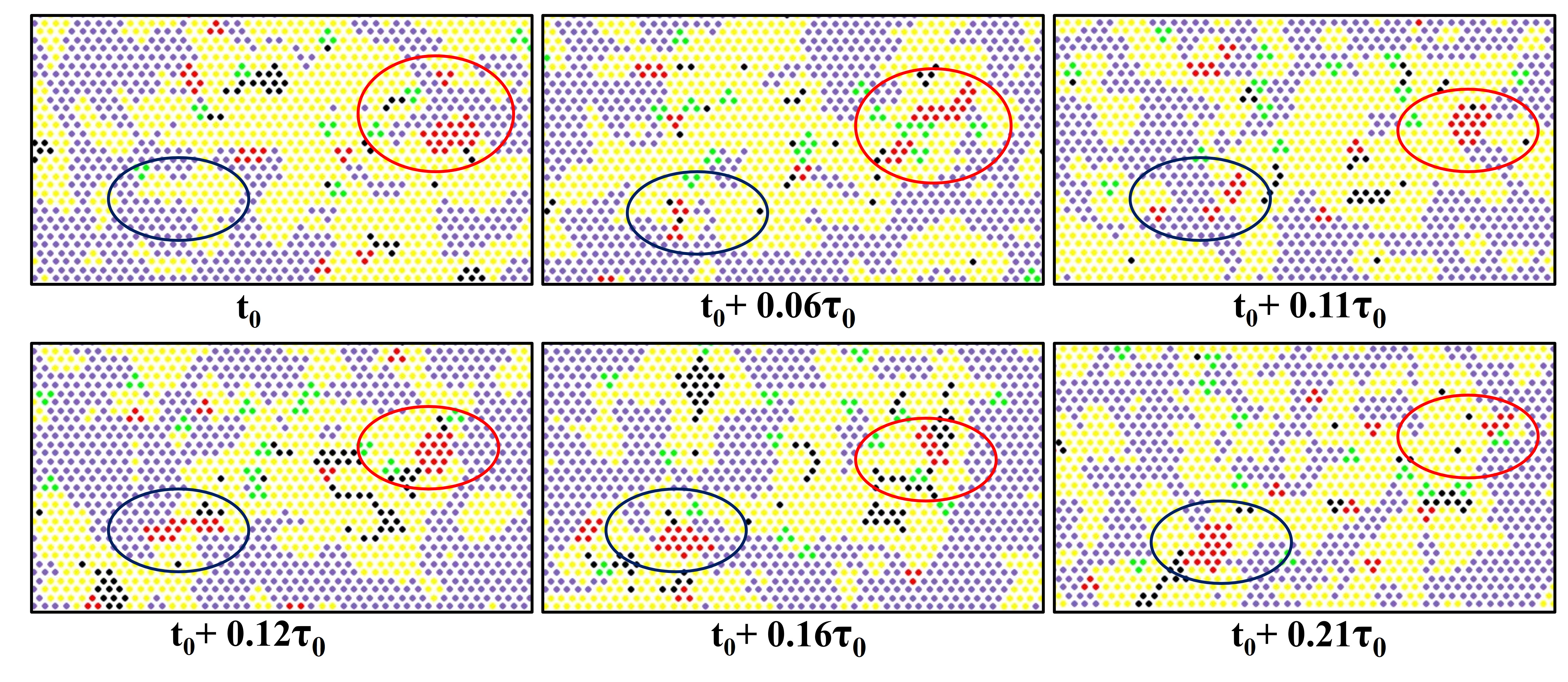}
   \centering
   \caption{Enlarged equilibrium snapshots illustrating the formation and dissolution of small protein clusters over time for switching rate $k = 10k_0$. The same color scheme as in Fig.~\ref{pic:kmcsnap} is used. The blue circle highlights a region where a small protein cluster forms, whereas the red circle highlights a region where a previously formed cluster dissolves. The total sequence spans a time interval on the order of the characteristic activation/deactivation period, $k^{-1}$. }
   \label{pic:10kclusterdyn}
\end{figure*}

The observed behavior arises from the coupling between protein diffusion in the membrane and the switching rate $k$. When $k$ is large ($10k_0$), the lifetime of the active state is short, and proteins typically switch to the inactive state before diffusing far enough to encounter other active proteins. As a result, active proteins form only small, transient aggregates (dimers/trimers), which rapidly dissolve due to frequent switching. This is illustrated in Fig.~\ref{pic:10kclusterdyn}, where small clusters appear and disappear on time scales of the typical activation time $\tau_0/10 =(10k_0)^{-1}$ (see also movie \texttt{movie\_10k0\_time.mp4}). In addition, the slower diffusion of small clusters compared to monomers further suppresses their coalescence. When $k$ is reduced ($k=k_0$), the active state persists long enough for proteins to diffuse farther and encounter one another more frequently, leading to the formation of larger and more stable oligomers [Fig.~\ref{pic:kmcsnap}(b); see also \texttt{movie\_k0\_time.mp4}].

For the smallest switching rate ($k=k_0/50$), active proteins can explore the entire system before deactivation, and dissolution of a large cluster would require the near-simultaneous deactivation of many proteins. Consequently, large and persistent aggregates dominate at small $k$ [Fig.~\ref{pic:kmcsnap}(c)].

    \label{tab:sysnam}
%\end{table*}

\section*{Conclusions}

In this work, we investigated membrane protein clustering in ternary lipid mixtures exhibiting $L_d$+$L_o$ coexistence using a previously developed lattice Monte Carlo model. We showed that the size and spatial organization of protein assemblies depend sensitively on two key parameters: the protein–lipid affinity $\varepsilon_{25}$ and the protein–protein interaction strength $\varepsilon_{55}$. The spatial distribution of proteins is determined mainly by their affinity for ordered lipids. When $\varepsilon_{25}$ is smaller than the DPPC–DPPC interaction energy ($\varepsilon_{22}=1.3$), proteins localize primarily in the $L_d$ phase or along extended $L_o/L_d$ interfaces. When $\varepsilon_{25}$ approaches $\varepsilon_{22}$, proteins partition more evenly between the two phases. For $\varepsilon_{25}$ comparable to or greater than $\varepsilon_{22}$, most proteins reside inside the $L_o$ domains, as shown in Fig.~\ref{pic:neib_stat}. In contrast, the extent of clustering is controlled by the relative magnitudes of $\varepsilon_{55}$ and $\varepsilon_{25}$. The resulting cluster size distributions (Fig.~\ref{pic:procluster}) show that weak protein–protein attraction ($\varepsilon_{55}<\varepsilon_{25}$) yields mostly monomers and small oligomers, comparable strengths ($\varepsilon_{55}\approx\varepsilon_{25}$) produce small stable clusters, and stronger interactions ($\varepsilon_{55}>\varepsilon_{25}$) lead to large, stable aggregates accompanied by the recruitment of saturated lipids. 

Increasing the protein concentration enhances this effect, promoting coarsening of the liquid-ordered domains, as demonstrated in Fig.~\ref{pic:pronumvar}. To put the conditions of our simulations in perspective, the 35DPPC mixtures contain roughly 6,500 lipids and 2,000 cholesterol molecules, so that a system with 100 proteins corresponds to a lipid-to-protein ratio of about 80–90, comparable to values reported for biological membranes. Despite this realistic molecular ratio, the total protein area coverage in our simulations is quite small (about 2\%), compared to 20–35\% typically observed in cell membranes. This is because the model proteins represent small inclusions, approximately three to four times smaller in diameter than typical transmembrane proteins. Introducing larger proteins would be computationally challenging, as their diffusion and state transitions are expected to suffer from high rejection rates, though efforts are currently underway to overcome these limitations.

To account for ligand-regulated binding and activation processes in biological membranes, we extended the model to a kinetic Monte Carlo framework in which proteins stochastically switch between inactive and active states with distinct interaction strengths. We find that the characteristic switching rate $k$, relative to the diffusion time $\tau_0$ required for a protein to traverse approximately one quarter of the lattice (comparable to the size of an $L_o$ domain), critically controls cluster size and stability. Rapid switching ($k \gg k_0=\tau_0^{-1}$) yields only transient dimers and trimers, while slow switching ($k \ll k_0$) reproduces the static limit with a large stable aggregate of active proteins. At intermediate rates, a broad distribution of cluster sizes appears. These results highlight how the interplay between molecular diffusion and activation lifetime may regulate the formation and persistence of membrane protein condensates, and may help rationalize ligand-dependent modulation of protein oligomerization observed in biological membranes.

In the present study, activation and deactivation rates were taken to be equal, whereas in biological systems these rates are often strongly asymmetric. Examining how unequal switching kinetics reshapes aggregation statistics will be an important next step. Likewise, we have treated all proteins as identical trimers with a single interaction pattern. Real membranes contain proteins of diverse sizes and interaction motifs, many of which can form both homo- and hetero-oligomers. Incorporating multiple protein species with distinct affinities, state-dependent interactions, and competition for lipid environments would allow the model to approach a wider range of biologically relevant behaviors.
Beyond these immediate extensions, the same lattice-based framework is well positioned to address more complex aspects of membrane organization that are central to biological function. Cellular membranes are asymmetric, with different lipid compositions in the two leaflets and transmembrane proteins spanning across them. Extending the model to a two-leaflet representation with controllable inter-leaflet coupling would enable systematic study of how such asymmetry influences phase behavior and protein localization. A further natural direction is to incorporate membrane curvature elasticity, for example by allowing vertical displacements of lattice sites and introducing phenomenological area and bending elastic terms. This would make it possible to examine how protein clustering and lipid demixing couple to membrane three-dimensional shape. Such an extension would allow investigation of curvature–composition feedback, domain-induced budding, and protein-stabilized scaffolds in a controlled and computationally tractable setting. Together, these developments would allow the lattice framework to bridge molecular interactions with mesoscale membrane organization, offering a route toward a deeper physical understanding of how lipid heterogeneity and protein assembly regulate biological membrane function.

\section*{Acknowledgments}
This work was supported by the Israel Science Foundation (ISF), Grant No. 1258/22.
\section*{Supplementary Material}
Online supplementary material contains MSD plot of proteins and trajectory movies.
%\clearpage
\bibliography{reference}
\end{document}